\newif\ifTR
\TRtrue
\ifTR
\documentclass[a4paper,USenglish]{lipics-tr}
\else
\documentclass[a4paper,USenglish]{lipics-v2016}
\fi
\usepackage[T1]{fontenc}
\usepackage[normalem]{ulem}
\usepackage[nounderscore]{syntax}
\usepackage{xspace}
\usepackage{lipsum}
\usepackage{graphics}
\usepackage{graphicx}
\usepackage{algorithmicx,algpseudocode,algorithm}
\usepackage{listings}
\usepackage{soul}
\usepackage{multirow}
\usepackage{multicol}
\usepackage{url}
\usepackage{breakurl}
\usepackage{float}
\usepackage{amsmath}
\usepackage{stmaryrd}
\usepackage[numbers,sort&compress]{natbib}
\usepackage{color}
\usepackage{xcolor}
\usepackage{todonotes}
\usepackage{amssymb}
\usepackage{amsthm}
\usepackage{booktabs}
\usepackage{csvsimple}

\usepackage{microtype}

\newif\ifBLIND
\BLINDfalse

\definecolor{lightgray}{rgb}{0.85,0.85,0.85}
\definecolor{darkgray}{rgb}{0.2,0.2,0.2}
\definecolor{purple}{rgb}{0.65, 0.12, 0.82}

\lstdefinelanguage{JavaScript}{%
  keywords={typeof, new, true, false, catch, function, return, null, catch, switch, var, if, in, while, do, else, case, break, throw},
  keywordstyle=\color{black}\bfseries,
  identifierstyle=\color{blue},
  sensitive=false,
  comment=[l]{//},
  morecomment=[s]{/*}{*/},
  commentstyle=\color{gray}\ttfamily,
  stringstyle=\color{darkgray}\ttfamily,
  morestring=[b]',
  escapechar=\%,
  morestring=[b]",
  escapeinside={/*@}{@*/}
}

\lstdefinelanguage{Trace}{%
  keywords={allocate,delete,null,undefined,true,false},
  keywordstyle=\color{black}\bfseries,
  identifierstyle=\color{black},
  sensitive=false,
  comment=[l]{//},
  morecomment=[s]{/*}{*/},
  commentstyle=\color{gray}\ttfamily,
  stringstyle=\color{darkgray}\ttfamily,
  morestring=[b]',
  morestring=[b]",
  escapechar=\%,
  escapeinside={/*@}{@*/},
  morestring=[s][\color{blue}\bfseries]{<}{>}
}

\newcommand{\appendixref}[1]{\autoref{#1}\xspace}
\newboolean{showcomments}
\setboolean{showcomments}{true}

\newboolean{showfill}
\setboolean{showfill}{true}
\newcommand{\Fill}[1]{\ifthenelse{\boolean{showfill}}{{\color{lightgray}\textit{\lipsum[1-#1]}}}{}}

\newcommand{\TL}{\textsc{TL}\xspace}
\newcommand{\TC}{Trace Collector\xspace}
\newcommand{\TT}{Trace Typer\xspace}

\newcommand{\TS}{\ensuremath{\mathbb{T}}\xspace}

\newcommand{\Ascribe}{\ensuremath{\alpha}\xspace}
\newcommand{\Generalize}{\ensuremath{\sqcup}\xspace}
\newcommand{\GeneralizeF}{\ensuremath{\overset{\rightarrow}{\sqcup}}\xspace}
\newcommand{\Merge}{\ensuremath{\mathtt{equiv}}\xspace}
\newcommand{\Typecheck}{\ensuremath{\mathtt{check}}\xspace}

\newcommand{\TAscribe}{\TS.\Ascribe}
\newcommand{\TGeneralize}{\TS.\Generalize}
\newcommand{\TGeneralizeF}{\TS.\GeneralizeF}
\newcommand{\TCheck}{\TS.\Typecheck}
\newcommand{\TMerge}{\TS.\Merge}

\newcommand{\TGenPoly}{\ensuremath{\TGeneralizeF_{\mathit{poly}}}\xspace}

\newcommand{\FlowIns}{\texttt{FI}\xspace}
\newcommand{\FlowSens}{\texttt{FS}\xspace}
\newcommand{\ContextIns}{\texttt{CI}\xspace}
\newcommand{\ContextSens}{\texttt{CS}\xspace}

\newcommand{\TypeEnv}{\ensuremath{\Gamma_{0}}\xspace}
\newcommand{\TypeEnvAbs}{\ensuremath{\hat{\Gamma}}\xspace}

\newcommand{\shapeMapN}{\ensuremath{\mathit{shapeMap}}\xspace}
\newcommand{\shapeMap}[1]{\ensuremath{\shapeMapN(#1)}\xspace}

\newcommand{\Var}[1]{\lstinline[language=Trace]!<#1>!\xspace}
\newcommand{\JS}[1]{\lstinline[language=JavaScript]!#1!\xspace}

\newcommand{\EvaluatedTypeSystemCount}{six\xspace}

\newcommand{\code}[1]{\lstinline{#1}}

\newcommand{\mypara}[1]{\vspace{-0.5em}\subparagraph*{#1}}

\newcommand{\trno}{{SRA-CSIC-2016-001}\xspace}

\ifTR
\title{Trace Typing: An Approach for Evaluating Retrofitted Type Systems (Extended Version)}
\else
\title{Trace Typing: An Approach for Evaluating Retrofitted Type Systems}
\fi

\ifBLIND
\author[1]{Authors suppressed for double-blind review}
\authorrunning{Suppressed for DBR}
\else
\author[1]{Esben Andreasen}
\author[2]{Colin S. Gordon}
\author[3]{Satish Chandra}
\author[3]{Manu Sridharan}
\author[3]{Frank Tip}
\author[4]{Koushik Sen}
\affil[1]{Aarhus University\\
  \texttt{esbena@cs.au.dk}}
\affil[2]{Drexel University\\
  \texttt{csgordon@cs.drexel.edu}}
\affil[3]{Samsung Research America\\
  \texttt{\{schandra,m.sridharan,ftip\}@samsung.com}}
\affil[4]{UC Berkeley\\
  \texttt{ksen@berkeley.edu}}
\authorrunning{E. Andreasen et al.} %
\fi

\Copyright{Esben Andreasen, Colin S. Gordon, Satish Chandra, Manu Sridharan, Frank Tip, Koushik Sen}%


\ifTR %
\EventNoEds{0}
\EventLongTitle{Samsung Research America Technical Report \trno}
\EventAcronym{SRA}
\EventYear{2016}
\EventDate{May 3, 2016}
\EventLocation{Mountain View, CA, USA}
\SeriesVolume{42}
\ArticleNo{23}
\else

\EventEditors{John Q. Open and Joan R. Acces}
\EventNoEds{2}
\EventLongTitle{42nd Conference on Very Important Topics (CVIT 2016)}
\EventShortTitle{CVIT 2016}
\EventAcronym{CVIT}
\EventYear{2016}
\EventDate{December 24--27, 2016}
\EventLocation{Little Whinging, United Kingdom}
\EventLogo{}
\SeriesVolume{42}
\ArticleNo{23}

\fi

\begin{document}

\maketitle

\begin{abstract}
   Recent years have seen growing interest in the retrofitting of type systems onto
   dynamically-typed programming languages, in order to improve type safety, programmer
   productivity, or performance.
   In such cases, type system developers must strike a delicate balance between
   disallowing certain coding patterns to keep the type system
   simple, or including them at the expense of
   additional complexity and effort.
   Thus far, the process for designing retrofitted type systems has been largely ad hoc,
   because evaluating multiple variations of a type system on large bodies of existing code
   is a significant undertaking.

   We present \textit{trace typing}: a framework for automatically and \textit{quantitatively} evaluating variations of
   a retrofitted type system on large code bases. The trace typing approach involves gathering
   traces of program executions, inferring types for instances of variables and expressions
   occurring in a trace, and merging types according to merge strategies that reflect
   specific (combinations of) choices in the source-level type system design space.

   We evaluated trace typing through several experiments.  We compared several variations of
   type systems retrofitted
   onto JavaScript, measuring the number of program locations with type errors in each
   case on a suite of over fifty thousand lines of JavaScript code.
   We also used trace typing to validate and guide the design of a new retrofitted type system that enforces fixed object layout for JavaScript objects.
   Finally, we leveraged the types computed by trace typing to automatically identify tag tests ---
   dynamic checks that refine a type --- and examined the variety of tests identified.
\end{abstract}

\def\sectionautorefname{Section}
\def\subsectionautorefname{Section}
\def\subsubsectionautorefname{Section}

\section{Introduction}\label{sec:introduction}
In recent years, there have been a number of efforts to \emph{retrofit}~\cite{lerner13} type systems onto dynamically-typed
languages, to aid developer productivity, correctness, and performance.  These languages are of increasing importance, primarily due to their common use in web
applications on both the client and server side.  As more large-scale, complex programs are written
in such languages, greater need arises for static types, due to the resulting benefits for
static error checking, developer tools, and performance.  Recent high-profile projects in type
system retrofitting include Closure~\cite{closure}, TypeScript~\cite{bierman14} and Flow for
JavaScript~\cite{flow}, and Hack for PHP~\cite{zhao12}.
Given the proliferation of dynamically-typed languages, there are many retrofitted type systems ahead.

The key questions in the design of these type systems involve selecting which features to include.
Richer features may allow more coding patterns to be understood and validated by the type checker.
However, richer type systems come at the cost of greater complexity and implementation effort, and
it is rarely a priori obvious whether a given feature's benefit outweighs its implementation
burden.
Hence, great care must be taken in deciding if a particular type system feature is worthy of
inclusion.  Such decision points continue to arise as a type system evolves: e.g., TypeScript
recently added union types~\cite{ts-unions}, and Flow recently added bounded
polymorphism~\cite{flow-bounds}.

When retrofitting a type system, there is generally a wealth of existing code to which one would like to introduce types.  When considering a type system feature, it would be very helpful to know \textit{how} useful the feature would be for typing an existing body of code.  Unfortunately, truly gauging the feature's usefulness would require implementing type checking (and possibly inference) for the feature and adding the necessary annotations to the code base. %
This involves a significant effort, which ultimately may not have lasting value if an alternate type system proves to be more suitable.
Some previous work~\cite{lerner13,papi08,dietl11,andreae06} reduces the first burden by providing a
reusable framework for implementing type systems, with some inference.  However, each still requires
manual annotation, so large-scale evaluation remains a substantial undertaking --- in one case
an expert required over 17 hours to annotate part of one program with type qualifiers~\cite{gordon13}.
For this reason, in
current practice, decisions are often made based on intuition and anecdotal evidence,
without the aid of carefully collected quantitative results on the impact of certain features.
Anecdotes and other qualitative criteria may validate inclusion of features that are rarely used but
important, but quantitative results on a large corpus can rapidly guide the high level design.

In this work, we propose a novel framework, called \textit{trace typing}, for \emph{automatic} and
\emph{quantitative} evaluation of retrofitted type system features for existing code bases.  Our
system works by collecting detailed, unrolled traces of dynamic executions of the code in question.
Evaluating a type system feature then requires implementing type checking and inference for the
unrolled trace, and also defining a \emph{merge strategy} (explained shortly) %
for combining the types of different occurrences of a variable in the trace into the corresponding static program entity.  The crucial observation here is that type checking and
effective inference for the unrolled trace is \textit{far} simpler than static type checking / inference for
the full source language, and
requires no type annotations due
to the simplicity of the trace format.  Hence, our system dramatically reduces the implementation
effort required to perform quantitative comparisons of different type system variants, and allows
type system designers to
\emph{automatically} gather feedback from large code bases by typing the trace.
\looseness=-1

While type inference and checking for unrolled traces is simpler to implement than a static type
checker, the ultimate goal of the designer is a type system for source programs, not traces.
Our key insight is that--given types for the unrolled trace--many source type systems can be
expressed via a family of \textit{merge} operations,
used to merge the types of runtime entities that the source type system cannot
distinguish.
For example, a merge operator can control whether assigning an integer and an object into the same variable yields a union type for the variable, or a type error.
(We
give examples of merging throughout the paper.)   The type errors found by using merged types
give an indicative lower bound on the set of program locations that would be
ill-typed under a static type checker, as long as the merge operator produces useful types. A
designer can, therefore, evaluate the usefulness of a proposed static type system via
suitable merge operators in the trace typing framework, and can also compare the relative power of
different static type systems.

We have implemented our techniques in a system for evaluating type system variants for JavaScript.
We evaluated our system on a diverse suite of programs,
and evaluated several different variations on a core object-oriented type
system.   In particular, our experiments yielded quantitative answers to the following questions for our benchmarks:
\begin{itemize}
\item How pervasive is the need for supporting union types on top of a type system that supports subtyping?
\item If union types are supported, what kind of ``tag tests'' usually occur to discriminate between branches of a union?
\item How pervasive is the need for supporting parametric polymorphism?
\item How pervasive is the need for intersection types for functions?
\end{itemize}
The results are given in \autoref{sec:experiments}. We believe trace typing
gives a quick and easy way to obtain answers to these (and similar) questions on a given code corpus.

We have also used trace typing to guide the design of a type
system to support ahead-of-time compilation of JavaScript extending previous work~\cite{choi15}.
Trace typing allowed us to validate some of our design choices \emph{without implementing a type inference engine} that works on real code. Moreover, it helped point to where we need to extend the type system.  Section~\ref{sec:sjs} details this case study.

Trace typing is designed to present a rapid, aggregate, quantitative picture to help a
type system designer decide whether to support certain features and
prioritize the development of a type checker.  When features are not supported,
a programmer needing said features must either ignore the warnings from the type
system, or refactor her code to fit the type system restrictions.
For example, if trace typing shows that parametric polymorphism is rarely needed in the code of interest, then the designer can defer supporting polymorphism; in the interim, developers must work around the limitation, e.g., by code cloning.
Of course, a type system
designer will need to consider other factors besides the output of trace typing when prioritizing, such as idioms used by critical modules, or the difficulty of converting code to fit within certain
restrictions.

\mypara{Contributions}
This paper makes the following contributions:
\begin{itemize}

\item We propose \emph{trace typing} as an approach for easily evaluating the usefulness of type
    system features, allowing large bodies of existing code to inform type system design.  Trace
    typing requires far less effort than implementing various static type checkers, and potentially
    adding type annotations to large bodies of code.

\item Using the trace typing approach, we systematically compare the precision of a number of object
type systems for JavaScript on some of the most popular packages for \texttt{node.js}, totaling
over 50,000 lines of code.  We include union types as well as several variants of polymorphism in our study.

\item We describe our experience with trace typing to guide and check choices in the design of a type
    system for use by an optimizing JavaScript compiler.

\item We use trace typing to automatically identify tag tests in our large corpus, and analyze the
    relative frequency of different tests and their structure.
\end{itemize}

\section{Trace Typing by Examples}\label{sec:overview}

In this section, we show informally how trace typing can be used to carry out different kinds of quantitative experiments relevant when designing a type system.  \autoref{sec:trace-typing} gives a more formal description of the framework.

\subsection{Polymorphism}
\label{sec:polymorphism-example}

\begin{figure}
\begin{lstlisting}[language=JavaScript]
function f(a) { a.p = 7; } /*@ \label{li:fun-f} @*/
function g(b) { return b; }
var x = { p: 3 }; /*@ \label{li:x-obj} @*/
var y = { p: 4, q: "hi" }; /*@ \label{li:y-obj} @*/
var z = { q: "bye", r: false }; /*@ \label{li:z-obj} @*/
f(x); /*@ \label{li:call-f-1} @*/// f.6: {p: Number} -> void
f(y); /*@ \label{li:call-f-2} @*/// f.7: {p: Number, q: String} -> void
g(y); /*@ \label{li:call-g-1} @*/// g.8: {p: Number, q: String} -> {p: Number, q: String}
var w = g(z); /*@ \label{li:call-g-2} @*/// g.9: {q: String, r: Boolean} -> {q: String, r: Boolean}
w.r = true; /*@ \label{li:deref-g-result} @*/
\end{lstlisting}
\caption{A JavaScript program to illustrate polymorphism.}
\label{fig:overview-example}
\end{figure}

\newcommand{\basets}{\ensuremath{\TS_{\emptyset}}\xspace}
\newcommand{\subts}{\ensuremath{\TS_{sub}}\xspace}
\newcommand{\boundts}{\ensuremath{\TS_{poly}}\xspace}
\newcommand{\fullts}{\ensuremath{\TS_{full}}\xspace}

Figure~\ref{fig:overview-example} gives a small JavaScript example.
The program
allocates new objects at lines~\ref{li:x-obj}--\ref{li:z-obj}, and then calls
functions \code{f} and \code{g}.  The \code{f} function accesses the \code{p} field, while \code{g} just returns its parameters.  All field accesses are well-behaved: they access
pre-existing fields, and fields are only updated with a value of a
compatible type.  Here, we show how trace typing can can \emph{quantify} the relative effectiveness of two type systems with different levels of polymorphism for handling this code.

We focus on the types observed
in the dynamic trace for functions \lstinline{f} and \lstinline{g}.
For our purpose, these
types correspond to the parameter and return types observed per
function call site.  So, for function \lstinline{f} we observe
the type \lstinline!{p: Number} -> void! for the call at line~\ref{li:call-f-1}, and
\lstinline!{p: Number, q: String} -> void! for line~\ref{li:call-f-2}.  Similarly,
for function \lstinline{g} we observe type \lstinline!{p: Number, q: String} -> {p: Number, q: String}! for the line~\ref{li:call-g-1} call, and
\lstinline!{q: String, r: Boolean} -> {q: String, r: Boolean}! for
line~\ref{li:call-g-2}.
Note that the types above are not based on inference on
the body of functions \code{f} and \code{g}.

Type system variants can be distinguished by how their merge
strategies combine these types into a single function type each for \lstinline{f} and for \lstinline{g}.  We consider the following merge strategies.
\subts{} merges, separately the argument and return types, in each case taking the least upper bound in a suitable subtyping lattice.
For our example, this strategy yields the type
\lstinline!{p: Number} -> void! for \lstinline{f} and
\lstinline!{q: String} -> {q: String}! for \lstinline{g}.\footnote{Note that
we did not compute a type for \lstinline{g} via least-upper-bound with respect to \textit{function} subtyping (with contravariance in argument position); see \autoref{sec:trace-type-system} for further details.}
\boundts{} merges the function types by introducing type variables, arriving at
a signature with parametric polymorphism (details in \autoref{sec:parametric-poly}).  The type we obtain for \code{f}
is the same as in the case of \subts, but for \code{g}, it is
\lstinline!(E) -> E!; \code{E} is a type parameter.

Next, the trace typing system
performs type checking using the merged types, and reports the
number of type errors to the user.  This type error count identifies
 program locations
that a static checker for the type system would be unable to validate, a valuable statistic for
determining the usefulness of the variant in practice (fewer errors\footnote{Throughout this paper,
we use \emph{error} as a synonym for \emph{static type error}, not to refer to actual developer mistakes in a
program.}
means that the variant can validate more code patterns).

With \subts, the types are sufficient for type-checking the bodies of the functions,
but \lstinline{g}'s type is still insufficient to typecheck the
dereference at line~\ref{li:deref-g-result}, leaving one error remaining.
With \boundts, the types are sufficient to ensure there is no type error in the program.
The type checking phase also checks conformance of the actual parameters being passed to function calls
at lines~\ref{li:call-f-1},
\ref{li:call-f-2},~\ref{li:call-g-1} and~\ref{li:call-g-2}; these checks pass for both \subts{} and \boundts{}.
\subts{} produces one type error on this trace, while \boundts{} produces 0.
These counts are underestimates with respect to a (hypothetical)
static type checker, both due to our reliance on
dynamic information, and due to the over-approximations inherent in any static type checking.  Nevertheless, across a large code base, such
data could be invaluable in finding the extent to which a particular
type system feature, e.g.\ parametric polymorphism, is useful.
Section~\ref{sec:subject-ts} presents results from an experiment that evaluated the usefulness of parametric polymorphism and a synthetic type system with ``unbounded'' polymorphism on a suite of benchmarks, via comparison of error counts.

\subsection{Discriminating Unions}\label{sec:tagged-union-example}

\begin{figure}
\begin{lstlisting}[language=JavaScript]
function f(x){
  ... x ... // x.2: A | int   /*@ \label{li:x-1} @*/
  if(x instanceof A){ // x.3: A | int  /*@ \label{li:x-2} @*/
    ... x ... // x.4: A  /*@ \label{li:x-3} @*/
  }
  ... x ... // x.6: A | int /*@ \label{li:x-4} @*/
}
f(new A()); /*@ \label{li:call-1} @*/
f(3); /*@ \label{li:call-2} @*/
\end{lstlisting}
\caption{Tag test example.}
\label{fig:tagtest}
\end{figure}

Suppose we wish to design a type system for JavaScript that is equipped to
reason about tag checks at the points where the program refines a union type.
Such a construct is shown in the example of Figure~\ref{fig:tagtest}.  In this example, a union type with more than one case is narrowed to a specific
case guarded by a condition.  In general, the guard can be an arbitrary expression, but that makes type checking difficult --- it would need to track an arbitrary number of predicates in carrying out an analysis.  For example,
the following ``non-local'' variation of the type guard in the example in Figure~\ref{fig:tagtest} would be more complex to handle:
\begin{lstlisting}[language=JavaScript]
  var y = x instanceof A
  ...
  if (y) { ...
\end{lstlisting}
It is simpler to support a limited number of syntactic forms.%

Can a type system designer obtain \emph{quantitative} feedback on what kind of
conditional constructs occur in the code corpora of interest?
Are non-local type guards common? What about conjunctions and disjunctions of predicates? Simple syntactic pattern matching is an
unreliable way to find this out, because it is fundamentally a flow analysis problem.

Trace typing makes this experiment easy.
Trace typing ascribes a type to each occurrence of each variable in the program based on dynamic
observation. %
Given these types, one can find pairs of successive reads of a variable without an intervening write, where the type
ascribed to the variable in the second read is different from that of the first. This is a heuristic to
identify instances of a variable's type becoming narrower on an unrolled path, likely to be a
tag test.

Consider the example in Figure~\ref{fig:tagtest} again.  We examine some of the type ascriptions carried out by trace typing during an execution of the program. Here, at line~\ref{li:x-1}, \code{x} refers to an object of type \code{A} and an
\code{int},
respectively, in the invocations from line~\ref{li:call-1} and line~\ref{li:call-2}.  Let us label the two dynamic occurrences of \code{x} as \code{x.2.1} and \code{x.2.2}, where the first index refers to the line on
which the variable occurs and the second refers to the invocation sequence
number.  The same types will be observed for \code{x.3.1} and \code{x.3.2}.
At line~\ref{li:x-3}, \code{x.4.1} observes \code{A}; that is the only dynamic
occurrence of \code{x} on this line.
Finally at line~\ref{li:x-4}, \code{x.6.1} and \code{x.6.2} observe
\code{A} and \code{int} respectively.

Crucially, trace typing can be configured to \emph{merge} the types ascribed to variable occurrences on
the same line, but across different invocations.  Say that our type system includes union types.  In that case, \code{x.2} (the result of merging \code{x.2.1} and \code{x.2.2}) gets the type
\code{A | int}, which represents a union type consisting of types \code{A} and \code{int}.  For \code{x.3} we get \code{A | int}
as well. However, for \code{x.4}, we get just \code{A}.   Hence, for successive
reads \code{x.3} and \code{x.4} (with no intervening write), we observe the situation that the second
type is a \textit{refinement} of the first.  This is a clue that the conditional at line~\ref{li:x-2} (that falls between the first and the second reads) is a type guard.

Using this technique on a large code corpus, we found that while non-local type guards are uncommon, boolean combinations of predicates are commonly used in type guards.  Section~\ref{sec:tagtests} details the experiment we carried out to find the nature of tag tests in real code.

\section{Trace typing}\label{sec:trace-typing}

\begin{figure}
  \centering
  \includegraphics[width=\textwidth,trim=0 370 0 30,clip]{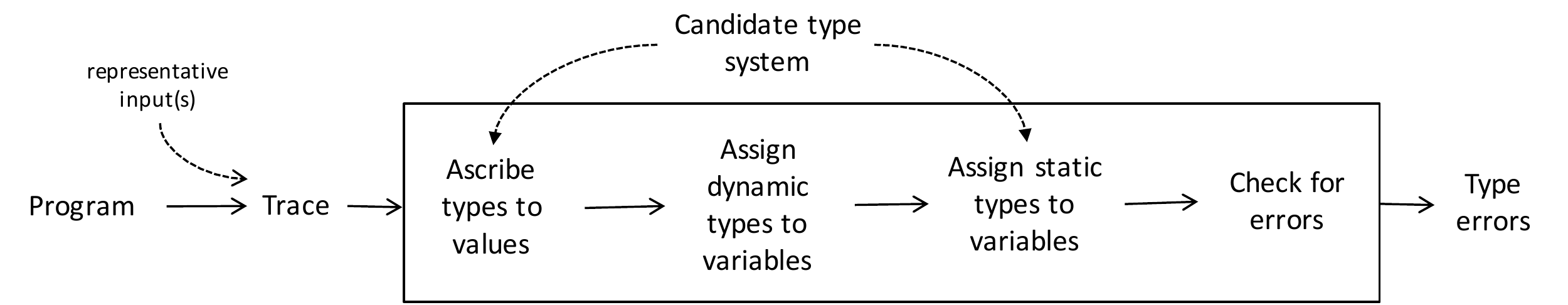}
  \caption{Trace typing framework overview.}\label{fig:framework-overview-drawing}
\end{figure}

\autoref{fig:framework-overview-drawing} gives an overview of the trace typing process. The trace typing framework takes two inputs:
\begin{itemize}
\item a type system of interest, specified by defining various parameters in a framework (\autoref{sec:trace-type-system}); and
\item a program \texttt{P} of interest, and inputs for exercising \texttt{P}.
\end{itemize}
Given these inputs, trace typing proceeds in the following four phases:
\begin{enumerate}
\item An unrolled \emph{trace} is produced by executing an instrumented version of \texttt{P} on the provided inputs (\autoref{sec:traces}).
\item The types of all values (including objects and functions) and variables in the trace are inferred (\autoref{sec:type-learning}).\footnote{In the remainder of the paper, we use the terms ``ascribed'' and ``inferred'' interchangeably.}
\item The inferred types are \emph{merged}, coarsening the precise types in the trace to a corresponding source typing (\autoref{sec:type-propagation}).
\item The resulting types are used for type checking, to estimate the ability of the type system to handle the observed behaviors (\autoref{sec:type-checking}).

\end{enumerate}

The number of errors reported in the last step is a useful metric for evaluating type system features (see \autoref{sec:discussion} for further discussion).  Our framework has proven to be sufficiently expressive to model even modern industrial proposals for retrofitted type systems (\autoref{sec:fixed-object-layout}).  Also, the computed types can
be used as part of experiments to test richer type system features, such as tag tests (\autoref{sec:detecting-tag-tests}).

\subsection{Trace Type Systems}\label{sec:trace-type-system}

Here, we describe the interface by which a type system is specified to the trace typing framework.
A type system \TS{} has five components:

\mypara{\TAscribe: (Value) $\rightarrow$ Type} ascribes a type to a concrete value.  For JavaScript, values include both primitives as well as object instances.

\mypara{\TGeneralize: (Type, Type) $\rightarrow$ Type} computes the least-upper-bound of the input types in the subtyping lattice for the type system.  For termination, we require \TGeneralize{} to be monotonic.

\mypara{\protect{} \TGeneralizeF{}: (Function-type\ldots) $\rightarrow$ Function-type} takes a list of types for individual invocations of a function \code{f} and computes a type that can accommodate all invocations.  A type for an invocation is computed by applying \TAscribe{} to the receiver, argument, and return values, and then combining them into a function type.
\TGeneralizeF{} often differs from applying \TGeneralize{} to the invocation types; further discussion below.

\mypara{\TMerge: (\FlowIns\ $\mid$ \FlowSens) $\times$ (\ContextIns\ $\mid$ \ContextSens)} governs how types for different trace occurrences of the same source variable are merged to approximate source-level typing (see \autoref{sec:type-propagation}).
The first component determines whether to maintain a separate type for a local variable at different program points (flow sensitivity), while the second component determines if separate types should be maintained across different function invocations (context sensitivity).  Note that use of \ContextIns{} vs. \ContextSens{} is usually dictated by the choice of \TGeneralizeF{} (see \autoref{sec:subject-ts}).  Also, our implementation supports more fine-grained flow- and context-sensitivity settings; here we only present two options for simplicity.

\mypara{\TCheck: (TypeEnv, Statement) $\rightarrow$ number} is a type checking function that
counts the number of type errors a trace statement causes in a given type environment.

Together, \TGeneralize, \TGeneralizeF, and \TMerge{} comprise the type system's \emph{merge strategy}.   \textbf{\textsf{Value}} represents the values of our trace language (see \autoref{sec:traces}), while
\textbf{\textsf{Type}} and \textbf{\textsf{Function-type}} are types
defined entirely by \TS{}.  Defining \TAscribe{} for JavaScript
objects is non-trivial, since the shape and property types of the
object may change over time; we detail the issues involved
in \autoref{sec:core-type-system}.

Note that \TGeneralizeF{} should \emph{not} use contravariance when handling function parameters, unlike typical function subtyping.  \TGeneralizeF{} is used to compute a type $\tau$ for function \code{f} that is consistent with all observed invocations of \code{f}.  Hence, each parameter type in $\tau$ should accommodate \emph{any} of the corresponding parameter types from the individual calls, a covariant handling.  For an example, see the handling of calls to \code{g} in \autoref{fig:overview-example} (\autoref{sec:polymorphism-example}).   %

\mypara{Example 1: Simple OO} To simulate a basic structural OO type system with width subtyping:
\begin{itemize}
    \item \TAscribe{} gives a type to an object by recursively characterizing the types of its
        properties (see \autoref{sec:core-type-system} for details).
    \item \TGeneralize{} gives the least-upper-bound of two object types (retaining common fields that have
	identical types) and returns the same type if given two identical types.
        Otherwise it returns $\top$.
    \item \TGeneralizeF{} applies \TGeneralize{} to the individual argument and return types (\emph{not} using standard function subtyping; see above) to compute a generalized function type.
    \item \TMerge{} is (\FlowIns,\ContextIns).
    \item \TCheck{} performs standard assignment compatibility checks for variable and field writes,
        property type lookup on property reads, and standard checks at function invocations.
\end{itemize}

\mypara{Example 2: Union Types}
To extend Example 1 with union types, we modify
\TGeneralize{} to introduce unions when merging different sorts of types (e.g., a number and object type).
For the code `\lstinline!var x = 3; x = { f : 4 };!', \texttt{x} would be given the union type $\textsf{number}\ \mid\ \{f:\textsf{number}\}$:
\autoref{sec:core-type-system} discusses how we perform type checking in the presence of such union types.

\mypara{Example 3: \subts{} and \boundts{} revisited}
We referred to \subts{} and \boundts{} in \autoref{sec:polymorphism-example}. \subts{} is precisely the Simple OO of Example 1. \boundts{} replaces \TGeneralizeF{} of \subts{} to introduce parametric polymorphism (described further in \autoref{sec:parametric-poly}).

\subsection{Traces}\label{sec:traces}
The execution of a source code program can be recorded as a finite sequence of primitive instructions that reproduces the dataflow of the execution. We call such a sequence a trace. A trace is a simplified version of the original source program:

\begin{itemize}
\item Every dynamic read or write of a variable or subexpression is given a unique identifier.

\item Control flow structures are erased: conditionals are reduced to the instructions of the conditional expression and chosen branch, loops are unrolled, and calls are inlined.

\item Native operations are recorded as their results; e.g.\ a call to the runtime's \code{cosine} function is recorded as the
    computed value.

\end{itemize}

We express our JavaScript traces as programs in a simple language \TL{}. \TL{} is an imperative,
object-oriented language without control flow structures or nested expressions. \TL{} has reads and writes for variables and fields, deletion for fields, and values (objects and primitive values).
The syntax of \TL{} can be seen in \autoref{fig:trace-language-grammar}.

\begin{figure}
    \small
    \begin{grammar}
      <trace> ::= <statement>*

      <statement> ::= <var> = <expression> "|" <var>.<name> = <var> "|" delete <var>.<name> "|" <meta>

      <expression> ::= <var> "|" <var>.<name> "|" "allocate" "|" "null" "|" "undefined" "|" <bool> "|" <number> "|" <string>

      <var> ::= "v1" "|" "v2" "|" "v3",  \ldots

      <name> ::= a "|" b "|" \ldots

      <meta> ::= "begin-call"<var><var><var>* "|" "end-call"<var> "|" "end-initialization"<var>

    \end{grammar}
  \caption{\TL{} grammar}
  \label{fig:trace-language-grammar}
\end{figure}

\begin{figure}[t]\scriptsize
\begin{lstlisting}[language=Trace,basicstyle=\scriptsize\ttfamily]
<tmp0@G> = allocate // ... allocation and assignment at line 6
<tmp1@G> = 3 /*@\label{line:tmp1_G-init}@*/
<tmp0@G>.p = <tmp1@G>
end-initialization <tmp0@1>
<x@G> = <tmp0@G> /*@\label{line:x_G-assign}@*/
begin-call <f@G> <x@G> // ... first call to f, at line 9
<a@f_1> = <x@G>
<tmp0@f_1> = 7
<a@f_1>.p = <tmp0@f_1>
end-call 
begin-call <f@G> <y@G> // ... second call to f, at line 10
<a@f_2> = <y@G>
<tmp0@f_2> = 7
<a@f_2>.p = <tmp0@f_2>
end-call
\end{lstlisting}
\caption{Excerpts from the trace for the program in \autoref{fig:overview-example}. The name of variables include their scope, e.g. \Var{x@G} is the \texttt{x} variable in the global scope, and \Var{a@f_2} is the \texttt{a} variable during the second call to \texttt{f}.}
\label{fig:running-example-trace}
\end{figure}

\autoref{fig:running-example-trace} shows excerpts from the trace for executing the code in \autoref{fig:overview-example}. Many JavaScript details are
omitted here for clearer exposition but handled by our implementation.
The variable names in the trace are generated from source variable names where present, and otherwise fresh temporaries are generated.  (Note that fresh temporaries are used within each call to \code{f}.)  A source mapping for each trace statement is maintained separately.

The trace contains some meta-statements to aid in later analyses.  Each call is demarcated by \texttt{begin-call} and \texttt{end-call} statements, to aid in recovering the relevant types for the invocation.  Data flow from parameter passing and return values is represented explicitly.  The \texttt{end-initialization} statement marks the end of a constructor or initialization of an object literal.

\subsection{Initial Type Ascription}
\label{sec:type-learning}

Given a trace and a type system as specified in \autoref{sec:trace-type-system}, trace typing first ascribes precise types to variables and values in the trace, without consideration for mapping types back to the source code.  First, types are ascribed to all values.  Primitive values and objects are handled directly using \TAscribe, while functions are handled by combining the type for each invocation with \TGeneralizeF, as described in \autoref{sec:trace-type-system}.

Once types for values have been ascribed, each trace variable is given
the type of the value it holds (recall that each trace variable is
only written once).  For a given trace $T$, we define the
initial type environment $\Gamma_0$ as: $\forall v\in\mathsf{Variables}(T)\ldotp \Gamma_0(v) = \TAscribe(\mathsf{ValueOf}(T,v))$.
This typing for trace variables is very precise---every variable access is given its own type, based solely on the value it holds at runtime.  The next phase
(\autoref{sec:type-propagation}) generalizes these initial types to more closely mimic the desired
source-level treatment of variables.

\mypara{Example} Consider the variable \Var{tmp1@G} at line~\ref{line:tmp1_G-init} in
\autoref{fig:running-example-trace}; it gets the value \JS{3}, for which $\subts.\Ascribe$ yields the
type \lstinline{Number}. Similarly the type of \Var{x@G} at line~\ref{line:x_G-assign} is
\lstinline!{p: Number}! because that is the type ascribed to the value in \Var{tmp0@G} at line~\ref{line:x_G-assign}.

\subsection{Type Merging and Propagation}\label{sec:type-propagation}

To model realistic type systems, we construct a less precise type environment \TypeEnvAbs{} by merging the types of ``equivalent'' trace variables in \TypeEnv.  Variable equivalence classes are determined by the \TMerge{} operator provided by the type system (see \autoref{sec:trace-type-system}).  If the first component of \TMerge{} is \FlowIns, we employ a flow-insensitive treatment of variables: all variable occurrences \texttt{v\_i} within a single dynamic call corresponding to the same source variable \texttt{v} are made equivalent.  No such equivalences are introduced if the first \TMerge{} component is \FlowSens.  If the second component of \TMerge{} is \ContextIns, we impose a context-insensitive treatment of variables: each matching \texttt{v\_i} \emph{across} function invocations (i.e., the same read or write of source variable \texttt{v}) is placed in the same equivalence class.

\begin{figure}
\begin{align*}
\TypeEnvAbs(x) & = \TypeEnvAbs_{m}(x)\ \TS.\sqcup\ \TypeEnvAbs_{p}(x) \quad\quad\quad\quad\quad\quad
\TypeEnvAbs_{m}(x) = {\underset{{x_{i}\in\hat{x}}}{\Large{\TS}.\bigsqcup}}\ \TypeEnvAbs(x_{i}) \\
\TypeEnvAbs_{p}(x) & = {\underset{'x = rhs'\in stmts}{\Large{\TS}.\bigsqcup}} \begin{cases}
  \TypeEnv(x), & rhs = \text{allocate} \vee rhs \in \textit{primitives}\\
  \TypeEnvAbs(y), & rhs = y\\
  \TypeEnvAbs(b).p, & rhs = b.p \wedge \TypeEnvAbs(b) \text{ is object with property } p\\
  \TypeEnv(b).p, & rhs = b.p \wedge (\TypeEnvAbs(b) \text{ not an object} \vee \TypeEnvAbs(b).p \text{ not present}) \\
\end{cases}
\end{align*}
\caption{Equations for type merging and propagation.}
\label{fig:scary}
\end{figure}

Given variable equivalence classes and $\Gamma_0$, \autoref{fig:scary} defines \TypeEnvAbs{} in terms of two components: $\TypeEnvAbs_{m}$ for \emph{merging} equivalent variables, and $\TypeEnvAbs_{p}$ for \emph{propagating} across assignments.  (As $\TypeEnvAbs_{m}$ and $\TypeEnvAbs_{p}$ are themselves defined in terms of \TypeEnvAbs, the equations must be solved by computing a fixed point.)  Given trace variable $x$, $\TypeEnvAbs_{m}(x)$ computes the least-upper bound ($\TS.\sqcup$) of all variables in its equivalence class $\hat{x}$.  However, this merging alone is insufficient for mimicking source typing, as it does not consider relationships between variables: if the program contains statement \code{x = y}, then the type of \code{y} influences the type of \code{x} in a source level type system.

$\TypeEnvAbs_{p}$ in \autoref{fig:scary} handles
type propagation across assignments.  The first case handles
assignments of values, using the baseline environment \TypeEnv.  For
the second case, $x = y$, we generalize $\TypeEnvAbs(x)$ to include
$\TypeEnvAbs(y)$.  This step is important to mimic source-level
handling of the assignment.  The final two cases handle object field
reads $b.p$.  If $\TypeEnvAbs(b)$ is an object type with a property
$p$, the handling is straightforward.  However, this may not hold, due
to other approximations introduced in computing $\TypeEnvAbs(b)$
(e.g., if $p$ were dropped when merging object types due to width
subtyping).  If $\TypeEnvAbs(b).p$ does not exist, we fall back to
$\TypeEnv(b).p$, i.e., we use the \emph{precise} type of $b$.  Without
this treatment, type errors could propagate throughout the remainder
of the trace, misleadingly inflating the type error count.  Our
fallback to the precise type of $b$ helps to localize type errors,
thereby allowing for gathering more useful information from the rest
of the trace than the alternative of producing $\top$.  For example,
if \lstinline!x! holds a number and later an object with property
\lstinline!foo!, \texttt{FI} merging will give it type $\top$.  A read
of \lstinline!x.foo! would produce a type error. Rather than tainting
all subsequent uses of the read result, we ascribe the type of the
value observed as the result of that read.

\mypara{Example} Consider type propagation for the \lstinline{a} variable in function \lstinline{f} of \autoref{fig:overview-example}, using type system \subts{} from \autoref{sec:overview}.  The corresponding variables in \autoref{fig:running-example-trace} are \Var{a@f_1} and \Var{a@f_2}, whose respective types are \lstinline!{p: Number}! and \lstinline!{p: Number, q: String}! in $\Gamma_0$ (\Var{y@G}'s initialization (line~\ref{li:y-obj} in \autoref{fig:overview-example}) is elided in \autoref{fig:running-example-trace}).  Hence, with the width-subtyping-based merge mechanism $\subts.\Generalize$, both \Var{a@f_1} and \Var{a@f_2} are assigned type \lstinline!{p: Number}! in \TypeEnvAbs.  The types in \TypeEnvAbs{} for all variables in \autoref{fig:overview-example} (using \subts) appear in \autoref{fig:var-types}.

\begin{figure}
  \centering
  \begin{lstlisting}[numbers=none]
 1: f :: <({p: Number}) -> Undefined>
 2: g :: <({q: String}) -> {q: String})>
 6: x :: {p: Number}
 7: y :: {p: Number, q: String}
 8: z :: {q: String, r: Boolean}
12: w :: {q: String}
  \end{lstlisting}
  \caption{Types ascribed for variable writes in \autoref{fig:overview-example} with \subts.}\label{fig:var-types}
\end{figure}

\subsection{Type Checking}
\label{sec:type-checking}

Once type propagation is done, type checking can be performed applying \TCheck{} to every statement
in the trace, using \TypeEnvAbs{} as the context.  \TCheck{} should return, for each statement, the number of falsified antecedents for the corresponding type rule (returning 0 implies a well-typed statement).

Note that our type propagation biases the location of type errors to occur more often at reads instead of writes.  Type propagation uses the \TGeneralize{} operator to find a variable type that handles all writes to that variable.  This process may produce a type higher in the subtyping lattice than what the programmer may have intended, causing type errors to be pushed to uses of that variable.

Consider the example: `\lstinline!x = {name: "Bob"}; x = {nam: "Bill"}; print(x.name);!'
Most likely, the programmer intended the type of \code{x} to include a \code{name} field.  If this
type were declared, the type checker would produce an error at statement two, due to the misspelled
\code{nam} field.  With trace typing under the \subts{} type system, we will instead compute a type
for \code{x} that has \textit{no} fields by merging the types of the two objects --- essentially
over-generalizing due to the presence of what is well described as a static type error.  Hence, the type error is pushed to the access of \code{name} at statement three, which is no longer in the type of \code{x}.

Intuitively, we want that when mimicking two comparable source-level type systems, trace typing ends up reporting fewer type errors for the more powerful one.  Indeed, our results show that this is generally the case.  However, this is not \emph{guaranteed} to be the case because our ascription operation (\TAscribe) can sometimes over-generalize, leading to unpredictable error counts.
See Section~\ref{sec:discussion} for further discussion of this matter.

\section{Instantiations}\label{sec:experiment-type-systems}

In this section, we detail several instantiations of the trace typing framework of \autoref{sec:trace-typing}, showing its flexibility.  We first describe a core type system that allows many of JavaScript's dynamic object behaviors.  We then show how this system can be used to detect possible tag tests in programs.  Then, we present an extension of this system to test the usefulness of parametric polymorphism.  Finally, we show an instantiation with a different, stricter type system for objects based on recent work~\cite{choi15}, and how to evaluate various features of that system using trace typing.

\subsection{Core Type System}\label{sec:core-type-system}

\begin{figure}
\newcommand{\dom}[1]{\mathsf{dom(#1)}}
\[
\begin{array}{rcl}
\tau & ::= & \top \mid \bot \mid \textsf{number} \mid \textsf{boolean} \mid \textsf{string} \mid
    \mathbb{O} \mid \mathbb{F} \mid \bigcup \overline{\tau} \mid \mu X.\tau\\
\mathbb{O} & ::= & \mathit{PropName}\rightarrow\tau\\
    \mathbb{F} & ::= & [\tau](\overline{\tau})\rightarrow\tau\mid \bigwedge \overline{\mathbb{F}}
\end{array}
\]
  \caption{Types used in the type systems we model.}\label{fig:type-system-grammar}
\end{figure}

The types we use for most of our experiments appear in \autoref{fig:type-system-grammar}.  The top
($\top$), bottom ($\bot$), primitive, union ($\bigcup$), and recursive types ($\mu$) are standard.
An object type ($\mathbb{O}$) is a map from property names to types.  Function types ($\mathbb{F}$)
are either standard (with receiver, argument, and return types), or an intersection type over
function types.  We use intersection types as a very precise model of call-return behavior, for a limit study of the usefulness of polymorphism (\autoref{sec:subject-ts}).

In our implementation, our object types are significantly more complex, mirroring the many roles objects play in JavaScript.  In JavaScript, functions are themselves objects, and hence can have properties in addition to being invokable.  Arrays are also objects with numeric properties and possibly non-numeric as well.  Our implemented object types model these semantics, but we elide them here for simplicity.  JavaScript also has prototype inheritance, where a read of property $p$ from object $o$ is delegated to $o$'s prototype if $p$ is not present on $o$.  For the core type system, object types collapse this prototype chain, so unshadowed properties from the prototype parent are collapsed into the child object type (more discussion shortly).  The type system of \autoref{sec:fixed-object-layout} deals with prototype inheritance more precisely.

\begin{figure}
\[\begin{array}{rcl}
  \shapeMap{o} & \equiv & \{ (p_i,v_i)\ |\ o.p_i = v_i \textrm{ at some trace point}\} \\
  \TAscribe(o) & \equiv & \{ (p_i,\tau_i)\ |\ M = \shapeMap{o} \wedge p_i \in \mathit{dom}(M) \wedge\ \tau_i = \underset{v \in M[p_i]}{\TGeneralize} \TAscribe(v) \}
\end{array}\]
\caption{Type ascription for objects.}
\label{fig:obj_ascription}
\end{figure}

\mypara{Ascribing Object Types}
As noted in \autoref{sec:trace-type-system}, defining the \TAscribe{} operation for object values is non-trivial.  Since object properties may point to other objects, types for object values are inter-dependent.  Further, JavaScript objects may be mutated in many ways: beyond changing property values, properties can also be added and deleted.  Even the prototype of an object may be mutated.

Trace typing is currently limited to object types that do not change over time, i.e., only one type can be associated with each object instance during the entire
execution.  So, for an object $o$, $\TAscribe(o)$ must merge together the different observed properties and property types for $o$ into a single object type.  Nearly all practical type systems use object types in this manner, as state-dependent object types require complex reasoning and restrictions around pointer aliasing.

\autoref{fig:obj_ascription} gives our definition of \TAscribe{} for objects.  We first define \shapeMap{o}, a map from property names to sets of values (shown as a set of pairs).  \shapeMap{o} captures all the property values that could be read from $o$ at any point in the trace.  (Note that we do not require the property read to actually exist in the trace.)  Since JavaScript property reads may be delegated to the prototype object, \shapeMap{o} gives a ``flattened'' view of $o$ that includes inherited properties.  Consider the following example:
\begin{lstlisting}[language=JavaScript]
var y = {};
var x = { a: 4 } proto y; /*@\label{li:map-ex-alloc}@*/ // prototype inheritance shorthand from /*@\color{gray}\cite{choi15}@*/ 
x.b = { c: false };
y.d = "hello";
x.a = 10;
\end{lstlisting}
For this example, the shape map for the object allocated on line~\ref{li:map-ex-alloc} is:
$[ \mathtt{a} \mapsto \{ 4, 10 \}, \mathtt{b} \mapsto \{ \mathtt{\{ c: false \}} \}, \mathtt{d} \mapsto \{ \mathtt{"hello"} \} ]$.

The object type $\TAscribe(o)$ is written as a set of pairs in \autoref{fig:obj_ascription}.  The properties in $\TAscribe(o)$ are exactly those in the domain of \shapeMap{o}.  The type for property $p_i$ is obtained by applying $\TAscribe$ to each value of $p_i$ in \shapeMap{o}, and then combining the resulting types using \TGeneralize.  When computing $\TAscribe$, cyclic data structures must be detected and handled by introducing a recursive type.

\mypara{Type Checking}
The type rules we checked in our implemented type systems are
standard.  Property accesses can be checked normally, since the
accessed property name is always evident in the trace.  Function
invocations are checked normally with one caveat: passing too
few arguments is permitted, as all JavaScript functions are implicitly
variadic.

For type systems with union types, we optimistically allow most operations that work for one case of the
union (e.g., property access on a union typed expression) without validating that the code has
checked for the appropriate case, but treat assignment of union-typed expressions to storage
locations soundly (TypeScript and Flow have very similar rules for unions).
     Our trace language does not currently include conditionals, making it non-trivial to use tag tests to eliminate union types~\cite{guha11,tobin-hochstadt10}.  However, our types do allow for discovering likely tag tests in the program, as we show below.

In the case when \TGeneralize{} causes a type to be $\top$, we count every use of and assignment to a variable of that type to be an error.

\subsection{Detecting Tag Tests}\label{sec:detecting-tag-tests}

To soundly eliminate union types, a type system must support narrowing the union under a conditional
that checks some tag information of a value.  While the theory for such narrowing is rich
and well-developed~\cite{guha11,tobin-hochstadt10}, industrial type systems for JavaScript (Flow and
TypeScript) treat union elimination unsoundly in the same ``optimistic'' manner we do
(\autoref{sec:core-type-system}).  A key issue is that tag tests can vary in complexity, from simple
type checks (e.g., JavaScript's \JS{typeof} operator) to complex combinations of user-defined
predicates, and it is unclear what level of complexity must be handled to provide adequate support
for tag tests.
\looseness=-1

Using trace typing, we designed a technique to observe which tag test constructs are being used most
frequently.  As described previously in \autoref{sec:tagged-union-example}, the technique works by
observing when there are two consecutive reads of a variable \code{v} (with no intervening write),
and \code{v}'s type is narrower at the second read than the first.  This technique requires using
the \FlowSens{} setting in \TMerge, to keep separate types for each variable occurrence.\footnote{Our
implementation also disabled the assignment merging shown in \autoref{fig:scary}, to discover more
tag tests.}  For function types, we used \TGeneralizeF{} from \subts, thereby generalizing using only
subtype polymorphism.  In \autoref{sec:tagtests}, we discuss the kinds of tag tests discovered in our benchmarks with this technique and their prevalence.

\subsection{Parametric Polymorphism}\label{sec:parametric-poly}

Here, we define \TGenPoly, an instantiation of the \TGeneralizeF{} operator that discovers types with parametric polymorphism.  With this approach, one can evaluate the usefulness of adding parametric polymorphism to a retrofitted type system.  %

\TGenPoly{} works by enumerating all possible type parameter replacements in each invocation type for a function, and finally choosing the most general parameterized type matching all invocations.  The enumeration proceeds by replacing all subsets of occurrences of a concrete type by a type variable.  Consider the identity function, \code{function id(x) \{ return x; \}}, and two invocations, \code{id(3)} and \code{id(true)}.  The observed invocation types for \code{id} are \lstinline! Number -> Number! and \lstinline!Boolean -> Boolean!.  (Applying \TGeneralizeF{} from \subts{} to these types would yield the unhelpful type $(\top)\rightarrow\top$ for \code{id}.)  For the first type, the enumeration would yield types \lstinline! X -> Number!, \lstinline! Number -> X!, and \lstinline!X -> X! (along with the original type).  After proceeding similarly for the second type, \TGenPoly{} returns the type \lstinline!X -> X! for \code{id}, as it is the most general type matching both invocations.

In general, \TGenPoly{} also generates signatures with multiple type parameters.
It also attempts to replace types nested one level into each argument or return type, to
discover types operating over generic data structures.  While this enumeration is exponential in the arity of the function, we did not observe an appreciable slowdown in practice.

After discovering polymorphic types, we must check the invocations against those types in \TCheck.  The checking can be done by treating each type variable as a direct subtype of $\top$, distinct from other type variables.  Our implementation currently uses an alternate strategy of modifying the context-sensitivity policy in \TMerge, so that variables in function invocations with distinct instantiations of the type parameters are not merged.  This strategy corresponds to checking a more powerful form of bounded polymorphism, which may yield a reduced error count compared to directly checking unbounded type parameters.

\subsection{Fixed Object Layout}\label{sec:fixed-object-layout}

We now show how to enhance the object types presented in \autoref{sec:core-type-system} to support \emph{fixed object layout}, as described in recent work by Choi et al.~\cite{choi15}.  In general, JavaScript objects is used as dictionaries, with arbitrary addition and deletion of properties (keys).  However, implementing all objects in this manner would lead to poor performance.
Real-world programs \emph{often} do not use this dictionary-like behavior.
Objects often behave like records, in which the set of properties is stable:
after initialization, new properties are not added or removed.  Modern just-in-time
compilers detect the record usage pattern at run time, and use it to optimize the object representation where possible.

Choi et al.~\cite{choi15} show how to enforce the record usage pattern in a type system, enabling efficient static compilation of a JavaScript subset.  Prototype inheritance complicates the design of such a type system, since a write of an inherited property creates a shadowing property in the child object.  Consider the following example:
\begin{lstlisting}[language=JavaScript,morekeywords={proto}]
var o1 = {a:3, m:function(x) {this.a = x;}}
var o2 = {b:5} proto o1 // prototype inheritance shorthand from /*@\color{gray}\cite{choi15}@*/ 
o2.a = o2.b; // adds field a to o2
\end{lstlisting}
The write at line 3 adds a local \code{a} property to \code{o2}, rather than updating the inherited property \code{o1.a}.  The type system for fixed layout handles this case by distinguishing between \textit{read-write} properties local to an object and \textit{read-only} properties inherited from the prototype chain.  With this distinction, the type system would reject the statement on line 3 above, as it is a write to a read-only property.\footnote{The type system of Choi et al.\ also tracks which properties are written on method receivers, to locally check validity of inheritance~\cite{choi15}.  With trace typing, this information need not be tracked explicitly, since receiver types are propagated into method calls.}

To track read-only and read-write properties in object types, we extend \TAscribe{} from \autoref{sec:core-type-system} as follows.  Instead of tracking a single shape map for each object, we keep a read-write shape map for locally-declared properties, and a read-only shape map for inherited properties.  In \autoref{sec:core-type-system}, the domain of \shapeMap{o} includes any $p_i$ present on $o$ at any point in the execution.  Here, to enforce the fixed layout property, we restrict the domain of each shape map to properties present before the end of $o$'s initialization.  (Recall from \autoref{sec:traces} that our traces include \texttt{end-initialization} statements to mark these points.)  Given these two shape maps, computation of read-only and read-write properties in each object type proceeds as in \autoref{fig:obj_ascription}.  Type checking is enhanced to ensure that property writes are only performed on read-write properties.

We also check for two additional properties from the Choi et al.\ type system~\cite{choi15}.  The type system requires any object used as a prototype parent to have a \emph{precise} type, i.e., no properties can have been erased from the type using width subtyping.  (This is required to handle subtle cases with inheritance~\cite{choi15}.)  In trace typing, we check this property by only flagging types of object literals as precise, so if width subtyping is ever applied, the resulting type is not flagged.  Finally, the type system requires that if a property from a prototype is shadowed, the parent and child properties have the same type.  This condition is easily checked in \TAscribe{} once read-only and read-write properties are computed.

\section{Implementation}
  \label{sec:implementation}

We implemented the trace typing process described in \autoref{sec:trace-typing}
as a tool%
for typing JavaScript programs.
This section reports on some of the more noteworthy aspects of the implementation.

\mypara{Tool Architecture}
Our tool consists of two components: \TC and \TT.
\TC is a Jalangi~\cite{sen13} analysis that obtains a trace of a
program of interest by monitoring the execution using source-level instrumentation; it consists of
about 2500 lines of JavaScript code.
\TT implements the core trace typing framework.  The framework and the type systems described in \autoref{sec:experiment-type-systems} are implemented in about 4000 lines of TypeScript, of
which about 1000 lines are the type system implementations.

\mypara{Modeling JavaScript Semantics}
In our implementation, great care is taken to accurately model JavaScript's complex primitive operators.  Much of the complexity in operator semantics lies in the implicit conversions performed on operands.  For example, the binary \lstinline!&&! operator first coerces each of its operands to a boolean value; this coercion can be applied to a value of any type.  However, the value of the \lstinline!&&! expression is always one of the \emph{uncoerced} operands.  The implementation creates trace statements that models these conversions explicitly, enabling accurate data flow tracking and type ascription.

\mypara{Modeling the Native Environment} To handle interactions
with the native environment (e.g., built-in JavaScript library
routines), we require a model of the native behavior.  We express
native models using trace statements, avoiding the need for a separate
modeling language.  In many cases, we can actually infer native models
from the concrete states of the execution, thereby avoiding the
significant work of writing models by hand.  Model inference works by
associating a new global \emph{escape variable} with each object that
can escape to the native environment.  Passing an object to the native
environment is modeled as a write to the corresponding escape
variable, while retrieving the object from the environment is modeled
as a read.  With this handling, a model can be automatically inferred
for much of the native environment, including complex functions such
as the overloaded Array constructor.  
\ifTR
Further details of this inference can be found in \appendixref{apdx:native_model}.
\else
Our technical report~\cite{tr} provides more details.
\fi

\section{Experiments}
\label{sec:experiments}

We report on several experiments we conducted using trace typing.  First, we studied
the trends in error counts for \EvaluatedTypeSystemCount{} type system variants, generated by varying
handling of subtyping and function types (\autoref{sec:subject-ts}), yielding useful insights on the
relative importance of these variants.  Second, we used trace typing to discover tag tests in our
benchmark and characterize their relative frequency and complexity (\autoref{sec:tagtests}).  Finally, we studied several questions around the restrictiveness
in practice of a type system for fixed object layout (\autoref{sec:sjs}).
Together, these experiments show the usefulness and versatility of the trace typing framework.  We
first describe our benchmarks, and then present the experiments.

\subsection{Benchmarks}\label{sec:benchmarks}

We use nine popular packages from \texttt{npm}, the node.js package manager, as benchmark programs.
\autoref{fig:benchmark-table} measures the static source code size of each 
program,\footnote{Non-comment, non-blank lines as counted by \texttt{cloc}, excluding build
scripts, tests and test harness code.} as well as the number of statements
in the traces we generate and the number of static source lines covered by our
executions.
\texttt{underscore} and \texttt{lazy.js} are utility libraries. %
\texttt{esprima} and \texttt{escodegen} are a JavaScript parser and code generator, respectively.
\texttt{typescript} is the compiler for the TypeScript language.\footnote{Figure
\ref{fig:benchmark-table} gives lines of JavaScript source (the compiler is
bootstrapped).}
\texttt{minimist}, \texttt{optparse} and \texttt{qs} are parsers for command line options and query strings.
\texttt{xml2js} is an XML serializer/deserializer.
These benchmarks capture a range of interesting JavaScript styles.  
The parsers, code generators, and compilers manipulate OO-style AST data structures
In contrast, the utility libraries and option parsers are highly reflective
(e.g., constructing or extending objects using dynamically-computed property names), but
are otherwise written in a largely functional style.  Hence, the suite as a whole exercises a variety of features of trace typing.
\looseness=-1

To exercise the programs for trace generation, we wrote a separate driver for each benchmark.  For the option parsers and \texttt{escodegen}, this driver comprised extracted tool invocations from their test suites. We exercised the utility libraries with code snippets from their official tutorials.

\begin{figure}
  \centering
\begin{tabular}{>{\bf}lcccccc}
&  \multicolumn{3}{c}{Benchmark sizes} & \multicolumn{3}{c}{Fixed-object-layout error rates} \\
\cmidrule(lr){2-4} \cmidrule(lr){5-7}
Benchmark               & \textbf{LOC} & \textbf{LOC exec} & \textbf{Length} & \textbf{ro/rw} & \textbf{prototypal} & \textbf{inheritance} \\
\cmidrule(lr){2-4} \cmidrule(lr){5-7}
\texttt{escodegen}  & 2132  & 723   & 375325 & 0/53    & 0/0  & 0/217 \\
\texttt{esprima}    & 4610  & 1052  & 23986  & 0/44    & 0/2  & 0/298 \\
\texttt{lazy.js}    & 2557  & 1016  & 25439  & 114/387 & 0/17 & 2/470 \\
\texttt{minimist}   & 186   & 151   & 140812 & 0/14    & 0/0  & 0/81 \\
\texttt{optparse}   & 222   & 141   & 15246  & 0/15    & 0/1  & 0/51 \\
\texttt{qs}         & 726   & 256   & 102637 & 0/41    & 0/0  & 0/62 \\
\texttt{typescript} & 35456 & 8139  & 167730 & 1/2085  & 2/3  & 0/3286 \\
\texttt{underscore} & 1098  & 727   & 120081 & 2/175   & 0/0  & 1/294 \\
\texttt{xml2js}     & 840   & 275   & 73672  & 0/42    & 1/2  & 0/91 \\
\cmidrule(lr){2-4} \cmidrule(lr){5-7}

Total               & 52193 & 14541 & 1243857 & 117/2856 & 3/25 & 3/4850\\
\end{tabular}
\caption{Benchmarks sizes and error rates for the fixed-object-layout
  type system. `LOC' is the lines of code in the program. `LOC exec'
  is the lines covered in our execution of the program. `Length'
  is the total number of trace statements. `ro/rw', `prototypal' and
  `inheritance' are the error-rates for different each kind of type
  check.}\label{fig:benchmark-table}
\end{figure}

\subsection{Comparing Type System Variants}\label{sec:subject-ts}

For our first experiment, we compared six type systems, generated by varying the \TGeneralize{} and \TGeneralizeF{} operators over the core types of \autoref{sec:core-type-system}.  For \TGeneralize, we used two options:
\begin{itemize}
\item \textbf{subtyping}: This uses standard structural (width) subtyping for objects, as in \subts{} from \autoref{sec:polymorphism-example} (see also Example 1 in \autoref{sec:trace-type-system}).
\item \textbf{unions}: \TGeneralize{} treats two object types as in \textbf{Subtyping}, but also introduces a union type when applied to an object type and a primitive (see Example 2 in \autoref{sec:trace-type-system}).
\end{itemize}
For \TGeneralizeF, we use three options:
\begin{itemize}
\item \textbf{base}: \TGeneralize{} is applied individually to argument and return types, as discussed in \autoref{sec:trace-type-system}.
\item \textbf{poly}: This uses \TGenPoly{} to discover parametric polymorphism, as discussed in \autoref{sec:parametric-poly}.
\item \textbf{intersect}: \TGeneralize{} simply combines two function types into an intersection of the two types.  This strategy yields more precise types than any practical type system, and hence is only used as a limit study of the usefulness of polymorphic function types.
\end{itemize}
\TMerge{} uses \FlowIns{} for flow-insensitive variable types in all configurations.  Context sensitivity is varied in accordance with \TGeneralizeF{}: for \textbf{base}, we use \ContextIns, while for \textbf{intersect} we use \ContextSens{} (since each invocation may have its own case in the intersection type).  Context sensitivity for \textbf{poly} was discussed in \autoref{sec:parametric-poly}.

After collecting traces, we ran our trace typer with each of our type system variants, counting the number of static (source location) type errors that arose.  We include only type errors that occur in the subject program, not in the libraries they depend on.

Experiments were run on a quad-core Intel Core i7-3520M@2.90GHz with 16GB RAM, running Ubuntu 15.10 and node.js 5.0.
Collecting traces and running all \EvaluatedTypeSystemCount{} configurations required less than 15
minutes.

\begin{figure*}[t]
  \centering
  \includegraphics[width=\textwidth,height=1.5in]{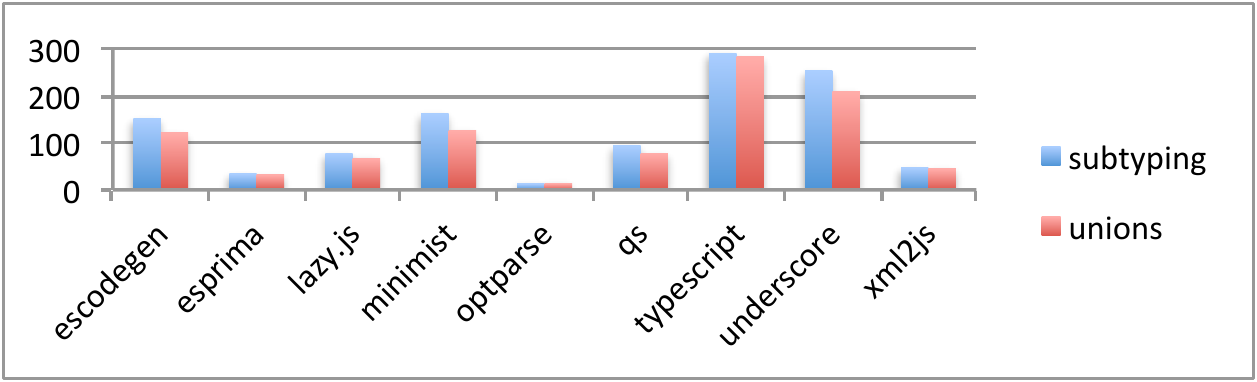}
  \caption{Error counts for \textbf{subtyping} and \textbf{unions}, selecting \textbf{base} for \TGeneralizeF.}\label{fig:by_base}
\end{figure*}

\begin{figure*}[t]
  \centering
  \includegraphics[width=\textwidth,height=1.5in]{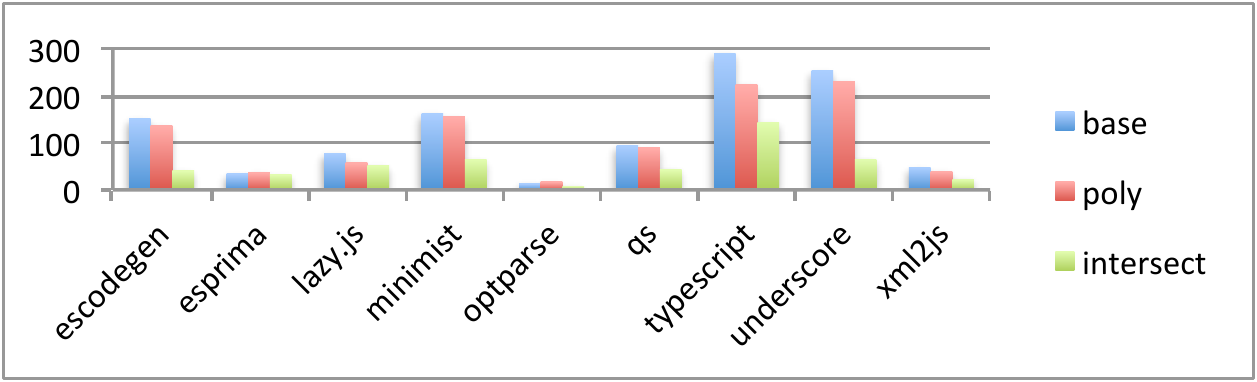}
  \caption{Error counts for different \TGeneralizeF{} settings, selecting \textbf{subtyping} for \TGeneralize.}\label{fig:by_sensitivity}
\end{figure*}

\ifTR
Our full data set is large---error counts for all configurations on all benchmarks---and is included in \appendixref{apdx:full-results}.
\else
Our full data set is available~\cite{tr}.
\fi
We present two select slices of data here.  Figure~\ref{fig:by_base} presents the number of syntactic locations with errors
for \textbf{subtyping} and \textbf{unions}, where \TGeneralizeF{} is \textbf{base} in either case.
Figure~\ref{fig:by_sensitivity} shows error counts for the three \TGeneralizeF{} settings, where \TGeneralize{} is \textbf{subtyping} in all three cases.  The trends shown here are similar across other configurations.

Overall we see trends we would expect, with more powerful type systems yielding fewer errors.
There are two small inversions in the results, where a more precise configuration (\textbf{poly})
produces more errors than a less precise configuration (\textbf{base}) discussed shortly.

This experiment let us find \emph{quantitative} answers to the kinds of questions we raised in \autoref{sec:introduction} and \autoref{sec:overview}.
We remind the reader that the objective of these experiments is not to pass verdict on the usefulness of certain type system features in an absolute sense. Rather the objective is to help designers of retrofitted type systems prioritize features based on empirical evidence rather than by intuition alone.
If a type system feature ends up unsupported, then users will need to
work around that limitation in their code (or live with type errors).
\looseness=-1


\mypara{How prevalent is the need for union types?} While \autoref{fig:by_base} shows that union types do reduce error counts over just using subtyping (as expected, since unions are more powerful), the degree of reduction is relatively small.
While TypeScript and Flow now both include unions, TypeScript did not add them until two years after its initial release.  The small error count reduction due to union types in our experiments could help to explain why they were not needed in TypeScript from the very beginning.\footnote{TypeScript's issue tracker has further discussion of their motivation for adding unions; see \url{https://github.com/Microsoft/TypeScript/issues/805}.}

\mypara{How prevalent is the need for parametric polymorphism? For intersection types?}
As shown in \autoref{fig:by_sensitivity}, the \textbf{intersection} configuration always reduced
errors compared to \textbf{poly}, while \textbf{poly} almost always reduces errors compared to
\textbf{base}\footnote{The inversions in \texttt{esprima} and \texttt{optparse} are due mostly to implementation quirks, such as an incomplete merge
operator on recursive types that returns $\top$ more than necessary.}.  While the error count reductions for \textbf{poly} are not dramatic in
general (\texttt{typescript} is discussed below), parametric polymorphism is crucial for ascribing types to core routines manipulating data structures (e.g., array routines), and hence its inclusion in TypeScript and Flow is unsurprising.
The dramatic drop in error count for several programs under the \textbf{intersection} configuration gives
strong evidence that JavaScript requires types that can express non-uniform polymorphism,
such as intersection types.  In fact, TypeScript has long supported overloaded functions for expressing intersection types.\footnote{See \url{http://www.typescriptlang.org/Handbook\#functions-overloads}.}

The \texttt{typescript} benchmark stands out from the others as it was originally implemented in TypeScript, and hence we would expect it to be mostly well typed.  Our results confirm
this: the error count is only 290 in 8139 (JavaScript) lines of code for the least precise
configuration, a much lower error rate than the other programs.  Also, the significant reduction in
errors going from \textbf{base} to \textbf{poly} for the benchmark (see
\autoref{fig:by_sensitivity}) corresponds nicely to the use of parametric polymorphism in the
TypeScript code base; see \autoref{sec:discussion} for an example.

While in the above experiment we used trace typing to retroactively find evidence  of the need of type system features, the next two experiments are concerned with more exploratory questions.

\subsection{Finding Tag Tests}\label{sec:tagtests}

Here, we provide the first empirical analysis of the use of tag tests in JavaScript which identifies
tag tests based on observed narrowing (as described in
Sections~\ref{sec:tagged-union-example} and~\ref{sec:detecting-tag-tests}), rather than syntactic criteria.
We post-process the ascription results to automatically identify occurrences of guard
predicates using the techniques of \autoref{sec:detecting-tag-tests}.
We then manually analyze the predicates to identify non-local guards (where the boolean
result of a check is computed outside a syntactic conditional statement), non-atomic guards (i.e.,
conjunction and disjunction of guards), and predicate functions (other functions whose result
indicates a certain refinement).

\mypara{Methodology} As mentioned in \autoref{sec:detecting-tag-tests}, we detect tag tests by observing consecutive reads of a variable where the type is narrowed at the second read.
Usually, the first read of this pair will be the use in the type guard.  In cases where the guard expression is saved to a local variable before being checked, this technique detected spurious additional guards, but they were easy to recognize and discard manually.

We ran the set of benchmark programs from \autoref{sec:benchmarks} with the union-producing
\TGeneralize operator of \autoref{sec:subject-ts}, and a flow sensitive \TMerge policy to keep types
of variable occurrences under conditionals separated.

\mypara{Results}
It took 90 minutes to manually classify all of the discovered
type guards, the results of which can be seen in \autoref{fig:classification}.  The figure contains rows for different kinds
of type guards, if a type is non-atomic, it can be classified as
having multiple kinds. Due to their prevalence, extra rows have been
added for type guards that are applied on the property of the guarded
value.

\begin{figure}
  \centering
  \begin{tabular}{lrl}
    \textbf{Measure} & \textbf{Count} & \textbf{Example} \\
    \midrule
    Type guards & 164 & - \\
    \midrule
    Non-local & 6 & \lstinline!var isObj = typeof x == 'object'; if(isObj)!\\ %
    Non-atomic & 46 & \lstinline$if(typeof x == 'string' || typeof x == 'object')$ \\
    \midrule
    typeof & 30 & \lstinline!typeof x === 'function'! \\
    instanceof & 9 & \lstinline!x instanceof Function !\\
    tag field& 16& \lstinline!x.kind = Kinds.Template!\\
   predicate function&30& \lstinline!Array.isArray(x)!\\
   property presence&35& \lstinline!x.hasOwnProperty('prop')!, \lstinline!x.prop === undefined!\\
   .prototype/.constructor&3& \lstinline!x.constructor === Function!\\
    \midrule
    .. test on property&37& \lstinline!typeof x.prop === 'function'!\\
    \midrule
    other&23&  \lstinline!className === toString.call(x)! \\
  \end{tabular}
  \caption{Tag test classification results.}
  \label{fig:classification}
\end{figure}

Overall there are 6 major classes of narrowing checks: \texttt{typeof}, \texttt{instanceof},
checking an explicit tag field in a data structure\footnote{A popular requested TypeScript
feature: \url{https://github.com/Microsoft/TypeScript/issues/186}} (mostly in the AST manipulations in our corpus),
a general predicate function, explicit property presence checks (\texttt{hasOwnProperty}), and checks of
objects' \texttt{prototype} or \texttt{constructor} fields.
The most prevalent type guards are \texttt{typeof}, use of general
predicate functions, and (a variety of) property presence tests.  Checking an explicit tag field, used to mimic
algebraic datatypes in JavaScript, are common but highly program-dependent.
Non-local type guards are rare, and would be easy to rewrite if needed.
Boolean combination of type guards is common enough that they probably should not be ignored by a
sound union elimination check.

\subsection{Evaluating Fixed Object Layout}
\label{sec:sjs}

We present a case study of evaluating
Choi et al.'s type system~\cite{choi15} to enforce \textit{fixed object layout} in JavaScript, as
described in \autoref{sec:fixed-object-layout}.
We checked the
following salient features in the trace typing framework:

\begin{enumerate}

\item \textit{Property writes} We checked that property writes were only performed on properties deemed to be ``read-write'' properties in the type system.

\item \textit{Precise types} We checked that only objects with a ``precise'' type (see \autoref{sec:fixed-object-layout}) were used as a prototype parent for other objects.

\item \textit{Consistent shadowing} We checked that when properties were shadowed in the prototype chain, the parent and child properties had the same type.

\end{enumerate}

We ran trace typing with this type system with the benchmark
suite of \autoref{sec:benchmarks} (with the subtyping \TMerge{} and \texttt{base} policy from
\autoref{sec:subject-ts}, merging all
occurrences of a variable in a dynamic function invocation).
\autoref{fig:benchmark-table} shows the errors reported by our trace typing model.
The column \textit{ro/rw} shows, in absolute numbers, how many of the property writes were to fields
that were inherited from prototype chain (and not shadowed by a local redefinition),
of the total number of property writes.  The column
\textit{prototypal} shows, how many of the assignments to some object's prototype property was
in violation of the restriction mentioned (second bullet) above, vs total number of such
assignments.  The column \textit{inheritance} shows in how many cases, a property was shadowed with
an inconsistent type.

The results add significant confidence to our initial
beliefs that some of the restrictions the type system imposes are not onerous in practice:
the numerator numbers are generally (though not universally) quite small compare to the denominator.
We manually found that most of the \textit{ro/rw} type errors for \texttt{lazy.js} were due to the way prototype hierarchies were initialized, and that the type errors could be eliminated by a simple refactoring.

Additionally, from the trace typing data (not presented here)
we found that support for
\emph{optional properties} in the type system could have helped eliminate a large number of type
errors, and so would be a very useful feature to add to the type system.
Trace typing finds this information without the need for implementing an actual static type inferencer
for JavaScript that is robust enough to work on large code bases.

\section{Discussion}
\label{sec:discussion}

In this section, we  discuss the quality of types inferred by trace typing,
present limitations and threats to validity, and then mention possible future generalizations of our approach.

\mypara{Quality of Types Inferred} Because our type inference approach is non-standard and our error counts are sensitive to over-generalization (Section~\ref{sec:type-checking}), it is worth asking whether the types we infer
are sensible --- whether a human developer would write similar types.  While there is no systematic
way to verify this, our experience says that the types are sensible.  In the course of developing the
framework and type system plugins, we spent a significant amount of type examining the types
inferred.  Object (property) types rarely deviated from what we would have written ourselves.

\begin{figure}
\begin{lstlisting}[numbers=none]
// TypeScript/src/compiler/parser.ts:39
export function forEachChild<T>(node:Node, ...):T {
  ... switch (node.kind) {
    ... case SyntaxKind.PropertySignature:
      ... return visitNode(cbNode, (<VarLikeDecl>node).propName);
    }
}
// TypeScript/src/compiler/type.ts:504
export interface VarLikeDecl extends Declaration {
  propName?: Identifier; \n dotDotDotToken?: Node; 
  name: DeclarationName; \n questionToken?: Node; 
  type?: TypeNode; \n initializer?: Expression;
}
\end{lstlisting}
\caption{Code fragment from \texttt{typescript}.}
\label{fig:tscode}
\end{figure}

We discuss one specific example of types inferred by trace typing for the \texttt{typescript} benchmark, for which the original TypeScript code contains declared types (trace typing analyzes the transpiled JavaScript version).  Figure~\ref{fig:tscode} contains an excerpt from the TypeScript code.
At the \lstinline!return! statement the \code{node} variable always has type \lstinline!VarLikeDecl! (also shown in the figure). Trace typing finds the relevant properties that we see mentioned in the interface, as well as the inherited properties.  Moreover, it also correctly computes the structural type of each property of \lstinline!VarLikeDecl! such as \lstinline!initializer!.
Due to space reasons, we do not show all the inferred types; but
we found them reasonable on manual inspection.

Our ascription of parametric polymorphic types (\autoref{sec:parametric-poly}) also works well in practice.  For example, it computes these
polymorphic signatures for array methods (from the native environment):
\begin{lstlisting}[numbers=none]
Array<E>.prototype.indexOf: (E) -> Number
Array<E>.prototype.concat: (Array<E>) -> Array<E>
Array<E>.prototype.push: (E) -> Number
Array<E>.prototype.pop: () -> E
\end{lstlisting}
These types are similar to the types used for native environment by TypeScript programs.\footnote{\url{https://github.com/Microsoft/TypeScript/blob/v1.7.3/lib/lib.core.d.ts\#L1008}}

The exceptions to the above arise mostly where type errors occur, and in those
cases the types are not bizarre but simply a different choice among several incompatible options
(e.g., when a method is overridden with an incompatible type).  For the \textbf{intersect} configuration of \autoref{sec:subject-ts}, with unbounded intersection types, the gap between inferred types and what one would write at the source level is larger, due to the synthetic nature of the type system.  Designing a trace typing system that infers a mix of parametric polymorphism and intersection types closer to what a person would write is future work.

\mypara{Limitations and Threats to Validity} Trace typing works under the assumption that the input programs are mostly type correct; the error locations generated by our framework lose accuracy when
this assumption is violated.  Moreover, we assume that most parts of a program \emph{have a correct
type} in the system being modeled, which may not always hold.
Our manual inspection of inferred types suggests that the natural formulations for \TAscribe{} and
\TGeneralize{} generally behave well, but it is possible other pathological cases exist.

A result that type system design $A$ yields fewer type
errors than design $B$ in our framework does \emph{not} strictly imply that the same will hold for
complete source-level implementations, due to inevitable over-approximations in a static type
checker.  Constructs inducing such approximations include reflective constructs like dynamic property accesses (which are resolved in our trace) and loops.
Type system feature interactions can exacerbate this effect.  The best use of trace
typing is to compare type systems that differ in only one or two parameters, and to give less
weight to very small differences in the number of errors detected.
Consider this example:
\begin{lstlisting}[language=JavaScript]
function f(x) { return x.p; }
f({p:3, q:4});
f({p:3, r:true});
var y = {p:3};
y = {p:3, q:5};
f(y);
\end{lstlisting}
Here, type ascription's over-generalizing in the presence of what would be static type
errors can lead to some unexpected results.  For the example above, ascription in the \textbf{base} polymorphism configuration (see \autoref{sec:subject-ts}) with no union types
would produce the type \lstinline!{p:Number} -> Number! for \code{f}, 
which would allow the trace of the program to type check.  Ascription with the more precise \textbf{intersect} configuration
would give the intersection type \lstinline!{p:Number,q:Number} -> Number! $\land$
\lstinline!{p:Number,r:Boolean} -> Number!.  But, the two writes to \texttt{y} ensures that it is ascribed type \lstinline!{p:Number}!, which is not a subtype of either argument type
in the intersection!  In such cases, a more powerful trace type system can produce more errors than
a less powerful one.  We did not observe this issue in experimental results we inspected manually, but it may have occurred elsewhere.
Our work is an empirical study using dynamic analysis, and we inherit the
normal threats to validity from such an approach, namely possible sampling bias in the
subject programs and sampling bias from the inputs used to gather traces.  For our study, we took inputs from each program's often-substantial test suite, which appeared from manual inspection to give reasonable path and value coverage.

\mypara{Potential Extensions} We have focused thus far on type systems for JavaScript, but the core ideas of our technique apply
to other languages as well.  The generalization to other dynamically typed languages (e.g., Ruby,
Python) would require instrumentation of the other language.  The type system implementations on top
of the new framework would need to be tailored to the primitives and idioms of the new target
language.

Our existing trace format could be used for additional experiments that analyze the types ascribed
to specific parts of the program.  For example, this could inform strategies for typing challenging constructs like
computed property accesses~\cite{politz12}.

More metadata could be added to our traces to enable experimentation with other type-system features, with no changes to our core approach.  For example, more detailed information about the evolving types and layouts of objects could be maintained, to enable experimentation with alias types~\cite{smith00,chugh12}.  Our hope is that trace typing will be a useful base for such future extensions.

\section{Related work}
\label{sec:related-work}

We focus here primarily on approaches to retrofitting type systems onto existing languages and on other means of evaluating type systems on large bodies of code, with a secondary focus on type systems for JavaScript.

\mypara{Retrofitting Type Systems}
Retrofitting type systems onto existing languages is not new,\footnote{The use of the term
\emph{retrofitting} is relatively recent, due to Lerner et al.~\cite{lerner13}.} nor is it restricted to dynamic
languages.  The notion of \emph{soft typing} for dynamically typed languages is now
well-established~\cite{cartwright91,wright97,tobin-hochstadt10}.
The notion of \emph{pluggable type
systems} --- including extensions to existing type systems --- is also an established
idea~\cite{bracha04}, which continues to produce useful tools~\cite{dietl11,papi08}.  Both
bodies of work suffer the same difficulty in evaluating lost precision: it requires tremendous
manual effort to evaluate on non-trivial amounts of code, leading to design decisions based on
intuition and a handful of carefully-chosen examples.

More recently, two groups of researchers arrived at means to induce a gradual type system variant
of a given base type system, one as a methodology to produce a gradual type system by careful
transformation of a static type system~\cite{DBLP:conf/popl/GarciaCT16} and another as a
a tool to compute a gradual version of a type system expressed as a logic
program~\cite{DBLP:conf/popl/GarciaCT16}.  Both deterministically produce one gradual type system
from one static type system, aiding with the introduction of gradual typing, but not of overall
design.

\mypara{Evaluating Type Systems}
The implementation effort for prototyping similar but slightly different type systems and applying
them is a significant barrier to evaluating completely new type systems for existing programming
languages.  \textsc{TeJaS}~\cite{lerner13} attempts to remedy this for JavaScript type systems by
carefully architecting a modular framework for implementing type systems.  It relies on
bidirectional typing to reduce annotation burden, but this still requires significant manual
annotation.

A smaller but closely-related body of work is evaluating changes to existing type systems.
Wright studied approximately 250,000 lines of SML code~\cite{wright95}, cataloguing the changes
necessary to existing code in order to type-check under the value-restriction to type
generalization.  The study in this case required less implementation effort than the general
approach we propose; SML implementations existed, and could be modified to test the single
alternative type system change.  Nonetheless, his careful analysis of the impact of the proposed
change justified a type system change that has withstood the test of time.  SML still uses the value
restriction, and OCaml uses a mild relaxation of it~\cite{garrigue04}.  More recently Greenman et
al.~\cite{greenman14} performed a similar evaluation of an alternative proposal for F-bounded
polymorphism in Java.

\mypara{Dynamic Type Inference}
Type inference for dynamic languages is another rich area with strong connections to our trace-based type inference.
The closest such work to ours is dynamic type inference for Ruby~\cite{an11}.  They use Ruby's
reflection capabilities to pass virtualized values through a program, with each use of the
virtualized objects gathering constraints on the type of the object.  Their system has a guarantee
that the inferred types will be sound if every path within each procedure is executed at least once.
Like trace typing, they rely on dynamic information to infer types for program fragments, including
for complex source-level constructs.  Unlike our approach, they make no attempt to simulate the
effect of type \emph{checking} the program: they infer types that hold at method entry and exit, but
do not infer types for local variables, whose types may change arbitrarily.  Also, they
cannot \emph{infer} method polymorphism (though they support annotations), require manual
annotation for native code, and evaluate only a single type system design.

Saftoiu et al.~\cite{saftoiu10} implement dynamic type inference for JavaScript, generating types
for a particular JavaScript type system based on dynamic observation, and find that it is useful for
converting code to their typed JavaScript dialect.  Their type system is sound with
respect to $\lambda_\textsf{JS}$~\cite{guha10}, and they can type-check the source code to verify the
inferred types. We cannot do this, because our goal is experimenting with type system design, so our
type systems are not adequate to type check source programs.  Like us they instrument code via
source-to-source translation. We evaluate significantly more code: their examples total 3799 lines of code.

\mypara{Other Uses of Traces} Coughlin et al.~\cite{coughlin12} use dynamic traces to identify the spans between program safety
checks and data uses guarded by those checks, e.g., a null check followed by a dereference.  These measurements can inform the design of static analyses, e.g., by indicating whether or not an
analysis should be interprocedural.  While similar in philosophy to our work, we differ
substantially in technical details.  First, their implementation is specific to null dereference
errors.  Second, their goal is to \emph{measure} aspects of program behavior quantitatively, and
interpret that measurement when designing an analysis; we instead advocate directly comparing
approximations of a real analysis.

{\AA}kerblom and Wrigstad~\cite{Akerblom:2015:MPP:2816707.2816717} study the receiver polymorphism of a very large base of open source Python
programs, primarily to examine how adequate nominal subtyping might be for Python (as opposed to
structural subtyping).  They classify static call sites by whether they were observed to have
multiple receiver types, consider whether call sites might be dispatched on
parametrically-polymorphic targets by clustering call sites in different ways (roughly similar to
our context-sensitivity policies), and do some reasoning about cases where different static calls
on the same syntactic receiver may have different qualities (a coarse approximation of typing local
variables).
They do not reason about actual method types, argument types (except implicitly if methods are
invoked on arguments), and do not attempt to simulate source type checking as we do.  ECMAScript 5
--- the previous version of JavaScript --- lacks classes, so even code written to ECMAScript 6 ---
which added classes --- will generally require structural subtyping.  Our type ascription machinery
could be used in a style similar to \autoref{sec:tagtests} to implement most of their
instrumentation.

\mypara{Type Systems for JavaScript}
Most previous work on static typing for JavaScript focuses on novel
approaches to typing its most expressive features; we describe particularly relevant
examples here. Guha et al.'s flow typing~\cite{guha11} allows for refining types in code
guarded by runtime tag checks. DJS~\cite{chugh12} implements a sophisticated dependent type system
that includes a type-based analogue~\cite{smith00} of separation logic to describe heap changes, and
an SMT-based refinement type system for characterizing the shape of objects. TeJaS~\cite{lerner13}
builds an expressive framework for experimenting with JavaScript type systems, whose core is an
extension to $F_{<:}^\omega$.

Throughout we have discussed Flow~\cite{flow} and TypeScript~\cite{ts-handbook,bierman14}.
They are quite similar (objects, unions with unsound elimination rules,
intersections) with a few small points of divergence (singleton string types and intentional
unsound subtyping in TypeScript, non-nullable fields in Flow).  They have very different approaches to
type inference and checking: TypeScript uses limited local type inference, while Flow performs a
global (per-module) data flow analysis.
We have already modeled key features --- objects with structural subtyping, union and
intersection types --- in our framework.

TypeScript and Flow include forms of parametric polymorphism, based on programmer annotations.
Trace typing can approximate some forms of parametric polymorphism (\autoref{sec:type-propagation}).
Both TypeScript and Flow also allow some narrowing of union types based on tag tests (unsound in general due to aliasing and mutation).  Trace typing could be extended to handle such features in future work, as discussed in \autoref{sec:discussion}.

Both TypeScript and Flow have added type system features based on user feedback, with great success.  Trace typing provides an alternate, complementary source of information for deciding which type system features to include, which can be employed without having a large user base.

\section{Conclusions}
  \label{sec:Conclusions}

We presented a framework for quantitatively evaluating variations of
a retrofitted type system on large code bases.  Our approach involves gathering
traces of program executions, inferring types for instances of variables and expressions
occurring in a trace, merging types according to merge strategies that reflect
options in the source-level type system design space, and type-checking the trace.
By leveraging the simple structure of traces, we dramatically reduce the effort required to get useful feedback on the
relative benefit of type system features.

To evaluate the framework, we considered \EvaluatedTypeSystemCount variations of a type system retrofitted onto JavaScript.
In each case, we measured the number of type errors reported for a set of traces
gathered from nine JavaScript applications (over 50KLOC) written in a variety of programming styles.  The data offers quantitative validation for
some of the design choices in two popular retrofitted type systems for JavaScript (Typescript and Flow).
In a different experiment, we used the results of trace typing to automatically identify places where type narrowing
occurred dynamically, to gather empirical results on the frequency and variety of tag tests in JavaScript.
In yet another experiment, we evaluated how onerous the restrictions for a new retrofitted JavaScript type system~\cite{choi15} were,
validating intuitions that our restrictions were mostly reasonable, and identifying priorities for future extensions.

The feasibility of carrying out these experiments is a strong validation of the trace typing approach.

\bibliographystyle{plainurl}
\bibliography{andreasen}

\ifTR
\appendix
\section{Modeling the Native Environment}
\label{apdx:native_model}

In Section~\ref{sec:implementation} we briefly mentioned that we automatically construct models of
the native environment.  This appendix gives full details.

We rely on three procedures, \texttt{toNative}, \texttt{fromNative} and
\texttt{allocate} (see \autoref{fig:nativeModel} for pseudo-code%
\footnote{
  Here, we only cover the important case where objects are exchanged with the native environment.
  Only minor changes to these steps are needed to account for primitives.
}) that performs this inference:
\begin{description}
\item[\texttt{toNative}:]
  This function models the four cases in which objects escape to native functions:
  the object is (i) the this-object or (ii) an argument in a call to a native function, or (iii)
  the object is returned to a native function as part of a callback or (iv) the object is thrown as an exception.
  An object, \lstinline!o!, that escapes to a native function forces a call to \lstinline!toNative(o)!, which
  uses a unique escape-variable for the object to ensure that the model represents the flow of that object
  regardless of how it is actually used in the native function.
\item[\texttt{fromNative}:]
  This function handles the four cases where objects enter the application from the native environment:
    the object is (i) the this-object or (ii) an argument in a callback from native code, or (iii)
    the object is returned from a native function, or (iv) the object is caught as an exception.
  An object, \lstinline!o!, that comes from the native environment forces a call to \lstinline!fromNative(o)!
  which will create a trace read-expression for the special escape-variable for the object. If the object has
  not been observed before, it assumed to be freshly allocated by the native environment, so a call to \lstinline!allocate! is also performed.
\item[\texttt{allocate}:] An object, \lstinline!o!, that is observed during the execution has an associated call to
    \lstinline!allocate(o)!,  which will create a trace allocate-expression and the necessary trace statements to
    initialize the fields and prototype of \lstinline!o!. This process might allocate further objects recursively.
    The initial global JavaScript object is modeled with a single call to \lstinline!allocate(global)!.
\end{description}

The use of an escape variable per escaping object \lstinline!o! may seem useless, but during type
propagation this results in the merge of each type at which the object \lstinline!o! escapes (e.g.,
if \lstinline!o! is thrown from source locations that see different subsets of its fields), and
allows that imprecision to manifest where the object re-enters the program from the native
environment (e.g., a catch of \lstinline!o! will see the merge of the types \lstinline!o! had at
throws) so we can soundly overapproximate the typing consequences of the native data flow.

As an example, consider the inference of a model for a call to the native Array constructor function.  This function takes a variable number of arguments, and returns an array containing those arguments; e.g., \lstinline!Array({p:1},{p:2})! returns \lstinline![{p:1},{p:2}]!.
The arguments to this call will be assigned to special escape variables via \lstinline!toNative!.  Then, a call to \lstinline!allocate! will
allocate the result array, as it is a fresh object from a native call.  Finally, via \lstinline!fromNative!, the elements of the
result array will be set to the corresponding arguments of the call by reading the corresponding special escape-variables. %

To handle cases where the automated inference described above would produce unsound results, we allow tool
users to specify that certain functions need to be modeled manually%
\footnote{
  For the experiments in this paper, we modeled the behavior of 11 functions manually.
}. In such cases, the user needs to create
a custom model that creates the appropriate trace statements.
This is the case for side effecting operations that may change the type of some property, such as \lstinline!Array.prototype.sort! which rearranges
the elements of an array. Note that, in such a rearrangement case, the model does not need to take care of the
general semantics of \lstinline!Array.prototype.sort!, just the concrete invocations: The trace statements that
reproduce the rearrangement can be generated by inspecting shape of the array before and after the call
to \lstinline!Array.prototype.sort!.

\begin{figure}
\begin{lstlisting}[language=,numbers=none]
toNative(o, fromVar) {
  makeTraceWrite(getEscapeVar(o), fromVar);
}

fromNative(o, toVar) {
  escapeVar = getEscapeVar(o);
  if(hasNotBeenAllocated(o)) {
    allocate(o, escapeVar);
  }
  makeTraceWrite(toVar, escapeVar);
}

allocate(o, allocVar) {
  makeTraceAllocation(allocVar)
  for(var field in o) {
    fieldVar = makeVariable();
    fromNative(o[field], fieldVar);
    makeTraceFieldWrite(allocVar, field, fieldVar);
  }
}
\end{lstlisting}
\caption{The three main procedures for inferring a model for the native environment.}
\label{fig:nativeModel}
\end{figure}

\section{Full Results for \EvaluatedTypeSystemCount{} Type Systems}
\label{apdx:full-results}

\begin{figure}
\begin{tabular}{lll}
\texttt{escodegen} & \texttt{esprima} & \texttt{lazy.js} \\
    \begin{filecontents*}{escodegen-clean.csv}
Base,Context,Err
subtyping,base,151
subtyping,poly,136
subtyping,intersect,40
unions,base,121
unions,poly,106
unions,intersect,25
\end{filecontents*}
\csvautotabular{escodegen-clean.csv} &
    \begin{filecontents*}{esprima-clean.csv}
Base,Context,Err
subtyping,base,34
subtyping,poly,36
subtyping,intersect,32
unions,base,32
unions,poly,34
unions,intersect,29
\end{filecontents*}
\csvautotabular{esprima-clean.csv} &
\begin{filecontents*}{lazy.js-clean.csv}
Base,Context,Err
subtyping,base,78
subtyping,poly,57
subtyping,intersect,51
unions,base,67
unions,poly,56
unions,intersect,51
\end{filecontents*}
\csvautotabular{lazy.js-clean.csv} \\
\texttt{minimist} & \texttt{optparse} & \texttt{qs} \\
\begin{filecontents*}{minimist-clean.csv}
Base,Context,Err
subtyping,base,162
subtyping,poly,155
subtyping,intersect,64
unions,base,126
unions,poly,122
unions,intersect,64
\end{filecontents*}
\csvautotabular{minimist-clean.csv} &
\begin{filecontents*}{optparse-clean.csv}
Base,Context,Err
subtyping,base,13
subtyping,poly,17
subtyping,intersect,8
unions,base,14
unions,poly,14
unions,intersect,9
\end{filecontents*}
\csvautotabular{optparse-clean.csv} &
\begin{filecontents*}{qs-clean.csv}
Base,Context,Err
subtyping,base,94
subtyping,poly,89
subtyping,intersect,43
unions,base,78
unions,poly,73
unions,intersect,40
\end{filecontents*}
\csvautotabular{qs-clean.csv}\\
\texttt{typescript} & \texttt{underscore} & \texttt{xml2js} \\
\begin{filecontents*}{typescript-clean.csv}
Base,Context,Err
subtyping,base,290
subtyping,poly,223
subtyping,intersect,144
unions,base,283
unions,poly,217
unions,intersect,128
\end{filecontents*}
\csvautotabular{typescript-clean.csv} &
\begin{filecontents*}{underscore-clean.csv}
Base,Context,Err
subtyping,base,253
subtyping,poly,231
subtyping,intersect,65
unions,base,208
unions,poly,190
unions,intersect,58
\end{filecontents*}
\csvautotabular{underscore-clean.csv} &
\begin{filecontents*}{xml2js-clean.csv}
Base,Context,Err
subtyping,base,47
subtyping,poly,39
subtyping,intersect,22
unions,base,45
unions,poly,39
unions,intersect,27
\end{filecontents*}
\csvautotabular{xml2js-clean.csv} \\
\end{tabular}
\caption{Trend experiment results comparing type systems with and without unions, for various models
of polymorphism.}
\label{fig:fullresults}
\end{figure}

\autoref{fig:fullresults} give the full results for the experiments in \autoref{sec:experiments}.

\fi

\end{document}